\documentclass[journal,twoside,web]{ieeecolor}
\usepackage{jsen}
\usepackage{cite}
\usepackage{amsmath,amssymb,amsfonts}
\usepackage{algorithm,algpseudocode}
\usepackage{graphicx}
\usepackage{textcomp}
\usepackage{wrapfig}
\usepackage{epstopdf}
\usepackage{comment}
\usepackage{bm}
\usepackage{paralist}

\newtheorem{remark}{Remark}

\def\BibTeX{{\rm B\kern-.05em{\sc i\kern-.025em b}\kern-.08em
    T\kern-.1667em\lower.7ex\hbox{E}\kern-.125emX}}
\definecolor{abstractbg}{rgb}{0.89804,0.94510,0.83137}
\begin{document}
\title{Outlier-Detection Based Robust Information Fusion for Networked Systems}
\author{Hongwei~Wang,~\IEEEmembership{ Member,~IEEE,}
        Hongbin~Li,~\IEEEmembership{Fellow,~IEEE,}
        Wei~Zhang,
        Junyi~Zuo,\\
        Heping~Wang
        and~Jun Fang,~\IEEEmembership{Senior Member,~IEEE}
\thanks{This work was supported in part by the China Scholarship Council and the National Natural Science Foundation of China (NSFC) under Grant 62103083, 11472222, and 61473227. The work of H. Li was supported in part by the National Science Foundation under Grants ECCS-1923739 and ECCS-2212940. This work is partly completed during the first author's visit to the Stevens Institute of Technology, Hoboken, NJ 07030.}
\thanks{H. Wang was with the School of Aeronautics, Northwestern Polytechnical University, Xi'an 710072, China; and is now with the National Key Laboratory of Science and Technology on Communications, University of Electronic Science and Technology of China, Chengdu 611731, China (e-mail: hongwei\_wang@uestc.edu.cn.com).}
\thanks{H. Li is with the Department of Electrical and Computer Engineering, Stevens Institute of Technology, Hoboken, NJ 07030 USA (e-mail: hli@stevens.edu).}
\thanks{W. Zhang, J. Zuo and H. Wang are with the School of Aeronautics, Northwestern Polytechnical University, Xi'an 710072, China (e-mail: weizhangxian@nwpu.edu.cn; junyizuo@nwpu.edu.cn; wangheping@nwpu.edu.cn).}
\thanks{J. Fang is with the National Key Laboratory of Science and Technology on Communications, University of Electronic Science and Technology of China, Chengdu 611731, China (e-mail: JunFang@uestc.edu.cn).}
\thanks{{\color{blue}{This paper is accepted by {\emph{IEEE Sensors Journal.}}}}} 
\thanks{{\color{red}{\copyright2022 IEEE. Personal use of this material is permitted. Permission from IEEE must be obtained for all other uses, in any current or future media, including reprinting/republishing this material for advertising or promotional purposes, creating new collective works, for resale or redistribution to servers or lists, or reuse of any copyrighted component of this work in other works.}}
}}

\IEEEtitleabstractindextext{%
\fcolorbox{abstractbg}{abstractbg}{%
\begin{minipage}{\textwidth}%
\begin{wrapfigure}[12]{r}{3in}%
\includegraphics[width=2.8in]{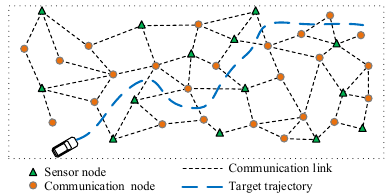}%
\end{wrapfigure}%
\begin{abstract}
  We consider state estimation for networked systems, where measurements from sensor nodes are contaminated by outliers. A new hierarchical measurement model is formulated for outlier detection by integrating an outlier-free measurement model with a binary indicator variable for each sensor. The binary indicator variable, which is assigned a beta-Bernoulli prior, is utilized to characterize if the sensor's measurement is nominal or an outlier. Based on the proposed outlier-detection measurement model, both centralized and decentralized information fusion filters are developed. Specifically, in the centralized approach, all measurements are sent to a fusion center where the state and outlier indicators are jointly estimated by employing the mean-field variational Bayesian inference in an iterative manner. In the decentralized approach, however, every node shares its information, including the prior and likelihood, only with its neighbors based on a hybrid consensus strategy. Then each node independently performs the estimation task based on its own and shared information. In addition, a distributed solution with approximation is proposed to reduce the local computational complexity and communication overhead. Simulation results reveal that the proposed algorithms are effective in dealing with outliers compared with several recent robust solutions.
\end{abstract}

\begin{IEEEkeywords}
networked systems, measurement outliers, outlier detection, centralized and decentralized information fusion, consensus, nonlinear information filter, variational Bayesian inference
\end{IEEEkeywords}
\end{minipage}}}

\maketitle

\section{Introduction}
\label{s1}

In recent years, networked systems (NSs) have attracted much attention with applications in various areas such as surveillance and patrolling, target tracking, intelligent transportation systems, and others~\cite{akyildiz2002survey,ferrari2010sensor,yu2018distributed,lin2020joint}. The growing interest in NSs has prompted intensive research on extending conventional state estimation methods for signal-sensor system, e.g., Kalman filter (KF)~\cite{kalman1960new} and its nonlinear variants~\cite{sarkka2013bayesian}, to cases involving NSs.

State estimation over NSs, in general, can be carried out in two main directions, i.e., the centralized approach and the decentralized one. In centralized solutions, all readings from sensors within the NS are transmitted to a fusion center that is responsible for processing the collected noisy measurements and providing state estimates~\cite{willner1976kalman}. The KF based solutions can be directly applied at the fusion center via a measurement-augmented approach. However, this incurs a high computational complexity due to the large dimension of the augmented measurements. To mitigate the computational burden, a variant of the KF, i.e., the information filter (IF), is frequently utilized. Several centralized state estimation solutions based on the IF were reported in~\cite{lee2008nonlinear,liu2012square,ge2016multisensor}.

In centralized solutions, data transmission may require significant communication overhead, thus constraining the scalability of NSs. Since the fusion center is the only signal processing unit, the NSs is critically dependent on the fusion center and would collapse when the latter fails. Furthermore, the fusion center must be provided with the knowledge of each sensor's measurement model and associated parameters, which creates additional challenges in estimation and communication, especially for heterogeneous NSs. In contrast, for the decentralized approach, each node within the NS has the ability to estimate the state using the information from itself and its neighbors via communication, either in multi-time-scale protocols~\cite{olfati2005distributed,carli2008distributed} or in single-time-scale protocols\cite{doostmohammadian2019cyber,
doostmohammadian2021distributed}. This allows decentralized systems to achieve a higher scalability and reliability. In addition, each node does not require the prior knowledge of the global network topology, making the decentralized approach suitable for time-variant NSs. Nonetheless, centralized solutions jointly process all observations within the network and thus offer more accurate estimation results than decentralized ones.

The decentralized approach needs to employ a proper strategy for information exchange among neighboring nodes in order to approach the performance of centralized solutions as much as possible. Consensus is a popular choice for this purpose. Several consensus strategies have been exploited for decentralized state estimation. A consensus on estimation (CE) strategy was proposed in~\cite{olfati2005consensus}, where the consensus was achieved by averaging local state estimates and predictions. Although easy to implement, the CE approach focuses on point estimation and ignores the error covariance which contains valuable information. To address this issue, an improved approach called consensus on measurement (CM) was proposed in~\cite{olfati2007}, which tries to make the local likelihood to reach an agreement, i.e., approximating the joint likelihood function in a distributed way. The convergence properties of the CM approach were examined in~\cite{kamgarpour2008convergence}, which shows that sufficient iterations are required to achieve the convergence of the the consensus procedure. Meanwhile, another approach, i.e., consensus on information (CI), was derived from the viewpoint of consensus on the local probability density functions (PDFs) in the Kullback-Leibler average sense~\cite{battistelli2014kullback}. The CI, unlike the CM, guarantees to converge to the local posterior PDFs with any number of consensus iterations (even only one iteration~\cite{battistelli2014kullback}). However, the information from measurements was overweighted as a result of its fusion rule. In~\cite{battistelli2014parallel} and \cite{battistelli2015consensus}, a consensus strategy called the hybrid CICM (HCICM) was proposed, based on an idea to integrate complementary features of the CI and CM, i.e., the stability guarantee with any number of consensus iterations of CI and the avoidance of any conservative assumption on the correlation when fusing the novel information of CM. The HCICM has been applied for distributed state estimation by integrating with the extended KF~\cite{battistelli2015consensus}, unscented IF (UIF)~\cite{battistelli2014parallel} and cubature IF (CIF)~\cite{chen2017distributed}.

The aforementioned state estimation methods, including both the centralized and decentralized, assume that measurement noises are Gaussian. In real applications, this assumption may not hold due to the presence of outliers. Several solutions have been proposed to deal with outliers. In~\cite{li2012distributed} the underlying non-Gaussian measurement noise was approximated by a Gaussian mixture model, and the CM strategy was utilized to develop a distributed information fusion algorithm. The interactive multiple model (IMM) approach was employed to develop robust solutions~\cite{battistelli2015,tian2015distributed}. To cope with the heavy-tailed measurement noise, a student's t distribution which can be interpreted as an infinity Gaussian mixture was used to fit the outlier-distributed Gaussian measurement noise, leading to a centralized state estimation algorithm in~\cite{zhu2013variational}, and a decentralized solution combined with the HCICM strategy in~\cite{dong2018robust}.

In this paper, we derive several robust state estimation solutions based on an outlier-detection strategy, using both the centralized and decentralized approaches. Outlier-detection has been of interest for many applications, and a multitude of solutions were developed in recent years, including, e.g., statistical solutions~\cite{wu2007localized}, distance based solutions~\cite{dai2020online}, classification based approaches~\cite{poornima2020anomaly}, and artificial intelligence based approaches~\cite{goodge2022lunar}. In our work, which aims to estimate states for a dynamic system in both centralized and decentralized manners, we utilize Bayesian solution for outlier detection. Specifically, a hierarchical measurement model is first introduced by integrating an outlier-free measurement model with a binary indicator variable that has a beta-Bernoulli prior. Based on the above model, a centralized information fusion solution is proposed, where the state and outlier indicators are jointly estimated by the mean-field variational Bayesian (VB) inference method. In the decentralized solutions, the VB method is utilized to estimate the state and indicator for each node, while the HCICM strategy is utilized to achieve consistency of all nodes. A target tracking example is studied to demonstrate the effectiveness of the proposed solutions.

The rest of this paper is organized as follows. Section~\ref{s2} formulates the problem of interest. The centralized and decentralized solutions are derived in Sections~\ref{s3} and~\ref{s4}, respectively. Section~\ref{s5} presents the numerical results and analyses. Conclusions are presented in Section~\ref{s6}.

\section{Problem Formulation}
\label{s2}

Consider a networked system with a set of nodes including communication nodes and sensor nodes which are distributed in a surveillance region. The topology of the network is modeled by an undirected graph $\mathcal{G}=(\mathcal{E},\mathcal{D})$, where $\mathcal{D}=\mathcal{S}\cup  \mathcal{C}=\{1,\cdots,N\}$ is the vertex set and $\mathcal{E}\subset\{\{i,j\}|i,j\in\mathcal{D},i\ne j\}$ is the edge set. $\mathcal{S}=\{1,\cdots,S\}$ is the set of sensors which have capabilities to make measurements. $\mathcal{C}=\mathcal{D}\backslash\mathcal{S}$ is the set of communication nodes which are used to improve the connectivity of the networked system. We assume that the network is connected, i.e., for any two vertices $i,j\in\mathcal{D}$, there exist a sequence of edges $\{i,a_1\},\{a_1,a_2\},\dots,\{a_k,j\}$ in $\mathcal{E}$. Let $N_{s}=\{j\in\mathcal{V}|\{s,j\}\in\mathcal{E}\}\cup\{s\}$ denote a subset that includes node $s$ and its neighbors.

The nonlinear discrete-time stochastic process observed by the networked system is described by the following state-space model:
\begin{align}
&\bm x_t=\bm f (\bm x_{t-1}) + \bm v_{t} \label{process}\\
&{\bm y}_{t,s}=\bm h_s (\bm x_t) + \bm w_{t,s},\quad s\in\mathcal{S}\label{measure}
\end{align}
where $\bm x_t\in\Re^n$ is the state vector; $\bm f(\cdot)$ is a known state evolution function; $\bm v_{t}\in\Re^n$ is the process noise, which is assumed to be Gaussian, i.e., $\mathcal{N}(0,\bm Q_{t})$; ${\bm y}_{t,s}\in\Re^{m_s}$ is a measurement made by the $s$-th sensor with respect to $\bm x_t$ at time instant $t$; $\bm h_s (\cdot)$ and $\bm w_{t,s}\in\Re^{m_s}$ are respectively, the measurement mapping and associated measurement noise of the $s$-th sensor; each measurement noise $\bm w_{t,s}$ is assumed to be Gaussian, i.e., $\mathcal{N}(0,\bm R_{t,s})$. The initial value of the state $\bm x_0$ is assumed to follow a Gaussian $\mathcal{N}(\hat{\bm x}_{0|0},\bm P_{0|0})$. In addition, the measurement noises of different sensor nodes are assumed to be independent of each other, and also independent with respect to the initial state and process noise.

The measurement model~\eqref{measure} is inadequate for some applications when the measurement may be contaminated by outliers. To account for potential outliers, we employ a binary latent variable $z_{t,s}$ as an indicator to characterize the state of the measurement $\bm y_{t,s}$. In particular, $z_{t,s}=1$ when $\bm y_{t,s}$ is a nominal measurement, while $z_{t,s}=0$ if $\bm y_{t,s}$ is an outlier. For Bayesian learning of the indicator variable, we impose a beta-Bernoulli prior~\cite{wang2018robust} on the indicator $z_{t,s}$. Therefore, the hierarchical model for measurement $\bm y_{t,s}$ in the presence of a potential outlier can be formulated as
\begin{align}
p(\bm y_{t,s}|\bm x_t,z_{t,s})&\propto\left(\mathcal{N}(\bm y_{t,s};\bm h_s (\bm x_t),\bm R_{t,s})\right)^{z_{t,s}}\label{likehood}\\
p(z_{t,s}|\pi_{t,s})&=\pi_{t,s}^{z_{t,s}}(1-\pi_{t,s})^{(1-z_{t,s})}\\
p(\pi_{t,s})&\propto{\pi_{t,s}^{e_{0,s}-1}(1-\pi_{t,s})^{f_{0,s}-1}}
\end{align}
where $\pi_{t,s}$ is a random parameter with $e_{0,s}$ and $f_{0,s}$ as its prior parameters to control the belief of $\bm y_{t,s}$ to be a nominal measurement or an outlier before the outlier detection procedure. In the proposed hierarchical model, $p(\bm y_{t,s}|\bm x_t,z_{t,s})$ is a standard Gaussian distribution when $z_{t,s}=1$, where it is a constant for the case where $z_{t,s}=0$. In the latter case, $\bm y_{t,s}$ can be effectively marked as an outlier since the likelihood is a constant and independent on the state. In general, the larger the value $e_{0,s}/(e_{0,s}+f_{0,s})$, the higher the probability that $\bm y_{t,s}$ is a nominal measurement.

The objective of this work is to develop solutions to estimate the states as well as to detect outliers for the networked system. We first present a solution based on centralized fusion. Although centralized fusion offers a performance benchmark, it has relatively low reliability and high communication overhead, as discussed in Section~\ref{s1}. We therefore also develop decentralized solutions, which perform state estimation in a decentralized manner. In this paper, we integrate consensus techniques with outlier detection for decentralized fusion.

\section{Centralized Robust CIF}
\label{s3}

For centralized processing, each sensor directly communicates with the fusion center. Specifically, each sensor sends its measurements to the fusion center where all the collected measurements are utilized to estimate the state. Then the fusion center feeds back the estimated state to each sensor if needed (as in a mobile sensor network where the sensor needs the state estimate to plan its trajectory). Since the measurements are mutually independent, the likelihood function of the measurements conditioned on all latent variables $\bm\Xi_t\triangleq\{\bm x_t, \mathcal{Z}_t,\bm{\pi}_t\}$ is given by
\begin{align}
p(\mathcal{Y}_t|\bm\Xi_t)=\prod_{s\in\mathcal{S}} p(\bm y_{t,s}|\bm x_t,z_{t,s})p(z_{t,s}|\pi_{t,s})p(\pi_{t,s})
\end{align}
where $\mathcal{Y}_t\triangleq \{\bm y_{t,1},\cdots,\bm y_{t,S}\}$, $\bm{\pi}_t\triangleq \{\pi_{t,1},\cdots,\pi_{t,S}\}$ and $\mathcal{Z}_t=\{z_{t,1},\cdots,z_{t,S}\}$. According to Bayes' theorem, the posterior distribution of all latent variables conditioned on $\mathcal{Y}_{1:t}$ is
\begin{align}
p(\bm\Xi_t|\mathcal{Y}_{1:t})=\frac{p(\bm\Xi_t,\mathcal{Y}_{1:t})}{p(\mathcal{Y}_{1:t})}
\end{align}
Due to the fact that $p(\mathcal{Y}_{1:t})$ is in general hard to calculate, obtaining the exact posterior distribution $p(\bm\Xi_t|\mathcal{Y}_{1:t})$ is computationally intractable. Therefore, some approximate methods should be employed. The variational Bayesian (VB) approach~\cite{Tzikas2008variational} is one such method, which uses a variational distribution $q(\bm\Xi_t)$ to approximate the posterior distribution $p(\bm\Xi_t|\mathcal{Y}_{1:t})$ by minimizing the Kullback-Leibler divergence (KLD) between $q(\bm\Xi_t)$ and $p(\bm\Xi_t|\mathcal{Y}_{1:t})$, i.e.,
\begin{align}
q(\bm\Xi_t) = \arg\min_{q}\text{KLD}\left(q(\bm\Xi_t)\|p(\bm\Xi_t|\mathcal{Y}_{1:t})\right)
\label{kld}
\end{align}
In this paper, we apply the mean-field approximation~\cite{Tzikas2008variational}, whereby the variational distribution is factorized as
\begin{align}
q(\bm\Xi_t)=q(\bm x_t)q(\mathcal{Z}_t)q(\bm{\pi}_t)\label{mfa}
\end{align}
Substituting~\eqref{mfa} into~\eqref{kld} and minimizing the KLD with respect to $q(\bm x_t)$, $q(\mathcal{Z}_t)$ and $q(\bm{\pi}_t)$ successively yields
\begin{align}
q(\bm x_t)&\propto\exp\Big(\langle\ln p(\mathcal{Y}_t,\bm\Xi_t|\mathcal{Y}_{1:t-1})
\rangle_{q(\mathcal{Z}_t)q(\bm{\pi}_t)}\Big)\label{vb1}\\
q(\mathcal{Z}_t)&\propto\exp\Big(\langle\ln p(\mathcal{Y}_t,\bm\Xi_t|\mathcal{Y}_{1:t-1})
\rangle_{q(\bm x_t)q(\bm{\pi}_t)}\Big)\label{vb2}\\
q(\bm \pi_t)&\propto\exp\Big(\langle \ln p(\mathcal{Y}_t,\bm\Xi_t|\mathcal{Y}_{1:t-1})
\rangle_{q(\bm x_t)q(\mathcal{Z}_t)}\Big)\label{vb3}
\end{align}
where $\langle g(\theta) \rangle_{q(\theta)}$ represents the expectation of $g(\theta)$ over the distribution of $q(\theta)$. It should be noted that $p(\mathcal{Y}_t,\bm\Xi_t|\mathcal{Y}_{1:t-1})$ is the full distribution of the SSM at time instant $t$, given by
\begin{align}
&p(\mathcal{Y}_t,\bm\Xi_t|\mathcal{Y}_{1:t-1})\notag\\
&\qquad=p(\bm x_t|\mathcal{Y}_{1:t-1})p(\mathcal{Y}_t|\bm x_t,\mathcal{Z}_t)p(\mathcal{Z}_t|\bm \pi_t)p(\bm\pi_t)
\end{align}
where $p(\bm x_t|\mathcal{Y}_{1:t-1})$ is the predictive density, which is approximated by a Gaussian distribution $\mathcal{N}(\hat{\bm x}_{t|t-1},\bm P_{t|t-1})$ given by~\eqref{pu_1} and~\eqref{pu_2}. Equations~\eqref{vb1}$-$\eqref{vb3} provide the update rules for the variational distributions, which are coupled. To address this issue, an alternating updating approach is generally employed in the VB inference, i.e., updating one variational distribution while fixing the others.

Computing the expectation in~\eqref{vb1} gives the following:
\begin{align}
q(\bm x_t)\propto \exp & \Big(-\frac{1}{2}\|\bm x_t-\hat{\bm x}_{t|t-1}\|_{\bm P_{t|t-1}^{-1}}^2\notag\\
&\qquad-\sum_{s\in\mathcal{S}}\frac{\langle z_{t,s}\rangle}{2}\|\bm y_{t,s}-\bm h_s (\bm x_t)\|_{{\bm R}_{t,s}^{-1}}^2\Big)
\label{up_x}
\end{align}
where $\|\bm x\|_{\bm A}^2={\bm x}^T{\bm A}{\bm x}$ and $\langle z_{t,s}\rangle$ is the mean of $z_{t,s}$. It is apparent that $q(\bm x_t)$ can be approximated by a Gaussian distribution $\mathcal{N}(\hat{\bm x}_{t|t},\bm P_{t|t})$ using the Kalman filtering framework, especially in its information format, for the multi-sensor data fusion problem. The parameter $\hat{\bm x}_{t|t}$ and $\bm P_{t|t}$ are obtained by
\begin{align}
\bm I_{t,s}&=\langle z_{t,s} \rangle \bm H_{t,s}\bm R_{t,s}^{-1}\bm H_{t,s}\label{I1}\\
\bm i_{t,s}&=\langle z_{t,s} \rangle \bm H_{t,s}\bm R_{t,s}^{-1}\tilde{\bm y}_{t,s}\label{i1}\\
\bm \Gamma_{t|t} &= \bm \Gamma_{t|t-1}+\sum_{s\in\mathcal{S}}\bm I_{t,s}\label{G1}\\
\bm \gamma_{t|t} &= \bm \gamma_{t|t-1}+\sum_{s\in\mathcal{S}}\bm i_{t,s}\label{g1}\\
\bm P_{t|t} & = \bm\Gamma_{t|t}^{-1}\label{P1}\\
\hat{\bm x}_{t|t}&=\bm P_{t|t}\bm\gamma_{t|t}\label{x1}
\end{align}
where $\bm \Gamma_{t|t-1}$ and $\bm \gamma_{t|t-1}$ are given by~\eqref{iv_1} and~\eqref{iv_2}, while $\bm H_{t,s}$ and $\tilde{\bm y}_{t,s}$ are calculated via~\eqref{p_H} and~\eqref{y_w} based on the different measurement mapping $\bm h_s(\bm x_t)$ and observation $\bm y_{t,s}$.

Since the components of $\mathcal{Z}_t$ are mutually independent, i.e., $p(\mathcal{Z}_t)=\prod_{s\in\mathcal{S}}p(z_{t,s})$, we can update them separately. For $q(z_{t,s})$, from~\eqref{vb2} we have
\begin{align}
q(z_{t,s})&\propto\exp \left \langle\ln p(\bm y_{t,s}|\bm x_t,z_{t,s}) + \ln p(z_{t,s}|\pi_{t,s})\right\rangle_{q(\bm x_t)q(\pi_{t,s})}\notag\\
&\propto\exp\Big\{-0.5z_{t,s}\text{tr}(\bm D_{t,s}\bm R_{t,s}^{-1})+z_{t,s}\zeta_1 \notag\\
&\qquad\qquad\quad\qquad\qquad\qquad\quad\quad+ (1-z_{t,s})\zeta_2\Big\}
\label{up_zz}
\end{align}
where
\begin{align}
&\bm D_{t,s}=\int{\left(\bm y_{t,s}-\bm h_s(\bm x_t)\right)\left(\bm y_{t,s}-\bm h_s(\bm x_t)\right)^T}q(\bm x_t)d\bm x_t\\
&\zeta_1\triangleq\langle\ln\pi_{t,s}\rangle_{q(\pi_{t,s})}=\Psi(e_{t,s})-\Psi(e_{t,s}+f_{t,s})\\
&\zeta_2\triangleq\langle\ln(1-\pi_{t,s})\rangle_{q(\pi_{t,s})}=\Psi(f_{t,s})-\Psi(e_{t,s}+f_{t,s})
\end{align}
with $\Psi(\cdot)$ denoting the digamma function. We can see from~\eqref{up_zz} that $z_{t,s}$ is a Bernoulli random variable with
\begin{align}
P(z_{t,s} = 1) &= c e^{\zeta_1-0.5\text{tr}(\bm D_{t,s}\bm R_{t,s}^{-1})}\\
P(z_{t,s} = 0) &= c e^{\zeta_2}
\end{align}
where $c$ is the normalized constant to ensure that $P(z_{t,s} = 1) + P(z_{t,s} = 0) = 1$. The expectation of $z_{t,s}$ is then given by
\begin{align}
\langle z_{t,s}\rangle_{q(z_{t,s})} = \frac{e^{\zeta_1-0.5\text{tr}(\bm D_{t,s}\bm R_{t,s}^{-1})}}{e^{\zeta_1-0.5\text{tr}(\bm D_{t,s}\bm R_{t,s}^{-1})} + e^{\zeta_2}}
\label{z}
\end{align}

Similarly, $q(\bm \pi_t)$ can be decomposed as $\prod_{s\in\mathcal{S}}q(\pi_{t,s})$ due to the independence. $q(\pi_{t,s})$ can be updated as follows:
\begin{align}
q(\pi_{t,s})\propto\exp\left((e_{t,s}-1)\ln\pi_{t,s}+(f_{t,s}-1)\ln(1-\pi_{t,s})\right)
\end{align}
with
\begin{align}
e_{t,s}&=e_{t,s}^0+\langle z_{t,s}\rangle_{q(z_{t,s})}\label{e}\\
f_{t,s}&=f_{t,s}^0+1-\langle z_{t,s}\rangle_{q(z_{t,s})}\label{f}
\end{align}
Clearly, $q(\pi_{t,s})$ is a Beta distribution $\text{Beta}(e_{t,s},f_{t,s})$.

For clarity, we summarize the centralized robust CIF (cRCIF) involving $K$-step VB iterations in Algorithm~\ref{alg1}.

\begin{algorithm}[!h]
  \caption{cRCIF} \label{alg1}
  \begin{algorithmic}[0]
    \State \textbf{Input:} $\mathcal{Y}_{1:T}$, $\hat{\bm x}_{0|0}$, $\bm P_{0|0}$, $\bm Q_{1:T}$, $\bm R_{1:T}$.
    \State \textbf{Output:} $\bm \hat{\bm x}_{t|t}$ and $\bm P_{t|t}$ for $t=1:T$.
    \For {$t=1:T$}
    \State Compute $\{\hat{\bm x}_{t|t-1},{\bm P}_{t|t-1}\}$ via \{\eqref{pu_1},\eqref{pu_2}\};
    \State Compute $\{{\bm \gamma}_{t|t-1},{\bm \Gamma}_{t|t-1}\}$ via \{\eqref{iv_1},\eqref{iv_2}\};
    \State Initialize $k=0$, $e_{t,s}^k$, $f_{t,s}^k$ and $\langle z_{t,s}^k\rangle=1$ for $s\in\mathcal{S}$;
     \For {$k = 1 : K$}
     \State Calculate $\{\bm I_{t,s}^k,\bm i_{t,s}^k\}$ via \{\eqref{I1},\eqref{i1}\} with $\langle z_{t,s}^{k-1}\rangle$;
     \State Update $\{\bm\Gamma_{t|t}^k,\bm \gamma_{t|t}^k\}$ via \{\eqref{G1},\eqref{g1}\};
     \State Update $\{\bm P_{t|t}^k,\hat{\bm x}_{t|t}^k\}$ via  \{\eqref{P1},\eqref{x1}\};
     \State Update $\langle z_{t,s}^{k}\rangle$ via \eqref{z} for $s\in\mathcal{S}$;
     \State Update $e_{t,s}^k$ and $f_{t,s}^k$ via \eqref{e} and \eqref{f} for $s\in\mathcal{S}$;
  \EndFor
  \State $\hat{\bm x}_{t|t} = \hat{\bm x}_{t|t}^K$, ${\bm P}_{t|t}={\bm P}_{t|t}^K$.
  \EndFor
  \end{algorithmic}
\end{algorithm}

\section{Consensus Based Decentralized Robust CIF}
\label{s4}

In this section, we derive two decentralized robust CIFs by integrating the hybrid-CICM consensus strategy with outlier detection. Note that in the decentralized solutions, outlier detection (i.e., VB iterations) is implemented at each sensor node, which is similar to the one in the centralized solution (in fact, both are identical when only one sensor is involved in the centralized solution). We therefore omit the details of the outlier detection procedure. In the following, we first briefly introduce the hybrid-CICM consensus strategy, and then explain how to integrate outlier detection with this consensus strategy to arrive at the first decentralized robust CIF (dRCIF-1). To further reduce both computational and communication burdens, we also propose an approximate implementation, referred to as the dRCIF-2. Some analyses of the proposed solutions are finally presented.

\subsection{Hybrid-CICM Consensus Strategy}

In this section, we provide a brief review of the hybrid-CICM consensus strategy. To facilitate description, we use the following operators:
\begin{align}
\bigoplus_{i}\left(\eta_i\odot p_i(x)\right)&\triangleq\frac{\prod_{i}(p_i(x))^{\eta_i}}{\int{\prod_{i}(p_i(x))^{\eta_i}dx}}\\
p_i(x)\oplus p_j(x)&\triangleq \frac{p_i(x)p_j(x)}{\int{p_i(x)p_j(x)dx}}
\end{align}
where $p_i(x)$ and $p_j(x)$ are some probability density functions (PDFs), and $\eta_i>0$ is a scalar. The consensus posterior density at the $s$-th node in the hybrid-CICM is given by~\cite{battistelli2014parallel,battistelli2015}
\begin{align}
p_{t,s}(\bm x_t)=p_{t|t-1,s}^L(\bm x_t)\oplus\left(\delta_{t,s}\odot r_{t,s}^L(\bm x_t)\right)
\label{con_post}
\end{align}
where $p_{t|t-1,s}^L(\bm x_t)$ is the result of consensus on prior, $r_{t,s}^L(\bm x_t)$ is the result of consensus on likelihood and $\delta_{t,s}$ is a weighting parameter to avoid overweighting on novel information. Clearly, obtaining $p_{t,s}(\bm x_t)$ requires three steps, i.e., consensus on prior, consensus on likelihoods, and fusing the consensus results of the priors and likelihoods (or the correction step in the Kalman filtering framework). In the following, we provide details to illustrate how to combine these three steps with the outlier detection procedure to obtain decentralized robust CIFs.

\subsection{Proposed Decentralized Robust CIF}

Since the local prior distribution is independent of the outlier detection procedure, the consensus on prior step can be carried out in the same approach as the conventional ones (e.g.,~\cite{battistelli2014paraller,battistelli2015}). Specifically, it can be obtained by the following $L$ iterations of the following averaging, i.e.,
\begin{align}
p_{t|t-1,s}^l(\bm x_t) = \bigoplus_{j\in{\mathcal{D}}}\Big(\kappa_{s,j}\odot p_{t|t-1,j}^{l-1}(\bm x_t)\Big)
\label{con_prior}
\end{align}
where $l=1,\cdots,L$, is the consensus step index and $\kappa_{s,j}$ is the consensus weight. In~\eqref{con_prior}, $p_{t|t-1,s}^0(\bm x_t)$ is initialized by $p_{t,s}(\bm x_t|\mathcal{Y}_{1:t-1})$. Since $p_{t,s}(\bm x_t|\mathcal{Y}_{1:t-1})$ is assumed to be Gaussian, the consensus on prior~\eqref{con_prior} has a closed form, with the precision and information vector updated by~\cite{battistelli2014kullback},
\begin{align}
\bm\Gamma_{t|t-1,s}^l=\sum_{j\in \mathcal{D}}\kappa_{s,j}\bm\Gamma_{t|t-1,j}^{l-1}\label{pr_c1}\\
\bm\gamma_{t|t-1,s}^l=\sum_{j\in \mathcal{D}}\kappa_{s,j}\bm\gamma_{t|t-1,j}^{l-1}\label{pr_c2}
\end{align}
The initialization parameters for~\eqref{pr_c1} and~\eqref{pr_c2} are $\bm\gamma_{t|t-1,j}^{0}=\bm\gamma_{t|t-1,j}$ and $\bm\Gamma_{t|t-1,j}^{0}=\bm\Gamma_{t|t-1,j}$. The consensus weight $\kappa_{s,j}$ is designed to satisfy $\kappa_{s,j}\ge 0$ and $\sum_{j\in N_j}\kappa_{s,j}=1$. In this work, we employ the Metropolis weights which are frequently used for consensus averaging~\cite{battistelli2014kullback}
\begin{align}
\kappa_{s,j}=\left\{\begin{array}{ll}
{\dfrac{1}{\text{max}\{|N_s|,|N_j|\}}}&s\in\mathcal{S},\ j\in N_s, j\ne s\\
 & \\
1-\sum\limits_{j\in N_s,n\ne j}\kappa_{n,j}& {j= s}\\
& \\
0 &\text{others}
\end{array}
\right.
\end{align}

\begin{algorithm*}[!t]
  \caption{decentralized robust CIF (dRCIF-1)} \label{alg_df2}
  \begin{algorithmic}[0]
    \State \textbf{Input:} $\mathcal{Y}_{1:T}$, $\hat{\bm x}_{0|0}$, $\bm P_{0|0}$, $\bm Q_{1:T}$, $\bm R_{1:T}$.
    \State \textbf{Output:} $\bm \hat{\bm x}_{t|t}$ and $\bm P_{t|t}$ for $t=1:T$.
    \For {$t=1:T$}
    \State For each node $s\in\mathcal{D}$, compute $\{\hat{\bm x}_{t|t-1,s},{\bm P}_{t|t-1,s}\}$ and $\{{\bm \gamma}_{t|t-1,s},{\bm \Gamma}_{t|t-1,s}\}$ via \{\eqref{pu_1}--\eqref{iv_2}\}.
    \State For each node $s\in\mathcal{S}$, compute the pseudo-measurement matrix $\bm H_{t,s}$ via~\eqref{p_H}.
    \For {$l=1:L$}
    \State For each node $s\in\mathcal{D}$, consensus on prior information $\bm\gamma_{t|t-1,s}^l$ and $\bm \Gamma_{t|t-1,s}^l$ \eqref{pr_c1} and \eqref{pr_c2};
    \EndFor
    \State For each node $s\in\mathcal{S}$, initialize $k=0$, $e_{t,s}^k$, $f_{t,s}^k$ and $\langle z_{t,s}^k\rangle=1$.
     \For {$k = 1 : K$}
     \State For each node $s\in\mathcal{S}$, calculate the local correction terms $\bm I_{t,s}^k$ and $\bm i_{t,s}^k$ via \eqref{I1} and \eqref{i1} with $\langle z_{t,s}^{k-1}\rangle$;
     \State For each node $s\in\mathcal{C}$, set the local correction terms as $\bm I_{t,s}^k=0$ and $\bm i_{t,s}^k=0$;
     \For {$l=1:L$}
     \State For each node $s\in\mathcal{D}$, consensus on the novel information $\bm I_{t,s}^{k,l}$ and $\bm i_{t,s}^{k,l}$ as \eqref{nv_c11} and \eqref{nv_c12};
     \EndFor
     \State For each node $s\in\mathcal{D}$, obtain the parameter $\kappa_t^s$ and update the total information $\bm \Gamma_{t|t,s}^k$ and $\bm\gamma_{t|t,s}^k$ via \eqref{f_c11} and \eqref{f_c12};
     \State For each node $s\in\mathcal{S}$, calculate $\bm P_{t|t,s}^k=(\bm \Gamma_{t|t,s}^k)^{-1}$ and $\bm x_{t|t,s}^k=\bm P_{t|t,s}^k\bm\gamma_{t|t,s}^k$, then update $\langle z_{t,s}^{k}\rangle$ via \eqref{z};
     \State For each node $s\in\mathcal{S}$, update $e_{t,s}^k$ and $f_{t,s}^k$ via \eqref{e} and \eqref{f};
  \EndFor
  \State For each node $s\in\mathcal{D}$, ${\bm P}_{t|t,s}=(\bm \Gamma_{t|t,s}^K)^{-1}$, $\hat{\bm x}_{t|t,s} = \bm P_{t|t,s}\bm \gamma_{t|t,s}^K$
  \EndFor
  \end{algorithmic}
\end{algorithm*}

Similarly, consensus on likelihood is performed by $L$-step iterations of the following:
\begin{align}
r_{t,s}^l(\bm x_t)=\bigoplus_{j\in{\mathcal{D}}}\Big(\kappa_{s,j}\odot l_{t,j}^{l-1}(\bm x_t)\Big)
\label{con_likelihood}
\end{align}
where $r_{t,s}^0(\bm x_t)$ is initialized as
\begin{align}
r_{t,s}^0(\bm x_t)=
\left\{
\begin{array}{ll}
p(\bm y_{t,s}|\bm x_t)& s\in\mathcal{S}\\
\text{constant}&s\in\mathcal{C}
\end{array}
\right.
\end{align}
Due to the presence of the indicator variable $z_{t,s}$, the initializing likelihood density $p(\bm y_{t,s}|\bm x_t)$ is no longer Gaussian. As a result, the consensus on likelihood~\eqref{con_likelihood} has no closed-form solution. Fortunately, the likelihood function conditioned on both $\bm x_t $ and $z_{t,s}$ is Gaussian, i.e., $p(\bm y_{t,s}|\bm x_t,z_{t,s})$ is a Gaussian distribution. Since the indicator variable $z_{t,s}$ is closely related to the VB iteration, the consensus on likelihood step is dependent on the VB iteration.

Given $z_{t,s}^k$ at the $k$-th VB iteration, the local likelihood of sensor node $s$ (i.e., $s\in \mathcal{S}$) can be approximated by
\begin{align}
p(\bm y_{t,s}|\bm x_t,z_{t,s}^k)\propto\exp\left(-\frac{1}{2}(\bm x_t^T\bm I_{t,s}^k\bm x_t-2\bm x_t^T\bm i_{t,s}^k)\right)
\end{align}
where $\bm I_{t,s}$ and $\bm i_{t,s}$ are respectively given by
\begin{align}
\bm I_{t,s}^k&= \langle z_{t,s}^k \rangle \bm H_{t,s}\bm R_{t,s}^{-1}\bm H_{t,s}\label{ll_1}\\
\bm i_{t,s}^k&=\langle z_{t,s}^k \rangle \bm H_{t,s}\bm R_{t,s}^{-1}\tilde{\bm y}_{t,s}
\end{align}
in which $\bm H_{t,s}$ and $\tilde{\bm y}_{t,s}$ can be found in~\eqref{p_H} and~\eqref{y_w}, respectively. For communication nodes (i.e., $s\in\mathcal{C}$), since the local likelihood is a constant, the information terms at the $k$-th VB iteration are
\begin{align}
\bm I_{t,s}^k =0, \ \bm i_{t,s}^k =0\label{ll_2}
\end{align}

Once the information terms related to the local likelihoods are obtained by~\eqref{ll_1}-\eqref{ll_2}, consensus on likelihood can be carried out by $L$ iterations of the following steps:
\begin{align}
\bm I_{t,s}^{k,l}=\sum_{j\in N_s}\kappa_{s,j}\bm I_{t,j}^{k,l-1}\label{nv_c11}\\
\bm i_{t,s}^{k,l}=\sum_{j\in N_s}\kappa_{s,j}\bm i_{t,j}^{k,l-1}\label{nv_c12}
\end{align}
with the following initialization
\begin{align*}
&\bm I_{t,s}^{k,0}=\bm I_{t,s}^k,\ \bm i_{t,s}^{k,0}=\bm i_{t,s}^k.
\end{align*}

After obtaining the consensus on prior and likelihoods, we then proceed to the correction step by fusing:
\begin{align}
\bm \Gamma_{t|t,s}^k&=\bm\Gamma_{t|t-1,s}^{k,L}+\delta_{t,s}\bm I_{t,s}^{k,L}\label{f_c11}\\
\bm \gamma_{t|t,s}^k&=\bm\gamma_{t|t-1,s}^{k,L}+\delta_{t,s}\bm i_{t,s}^{k,L}\label{f_c12}
\end{align}
where $\delta_{t,s}$ is a scale parameter used to avoid overweighting the novel information. In principle, a reasonable selection of $\delta_{t,s}$ is $|\mathcal{N}|$ since the consensus weight $\kappa_{t,s}^L=1/|\mathcal{N}|$ when $L\to\infty$, and such a choice makes the distributed filter converge to a centralized one when the consensus iteration tends to infinity~\cite{battistelli2015}. In practice, however, the number of consensus iterations is small due to the constraint of power supply of each node, creating some problem with the choice of $\delta_{t,s}=|\mathcal{N}|$, as shown in~\cite{battistelli2015}. An alternative is to compute $\delta_{t,s}$ in a distributed approach, i.e.,
\begin{align}
\delta_{t,s} = \left\{\begin{array}{ll}
1,&\theta_{t,s}^L=0\\
1/\theta_{t,s}^L, &\text{else}
\end{array}
\right.
\label{delta}
\end{align}
where $\theta_{t,s}^L$ is iteratively determined via
\begin{align}
\theta_{t,s}^l=\sum_{j\in N_s}\kappa_{s,j}\theta_{t,j}^{l-1}
\end{align}
with $\theta_{t,s}^0=1$ if $s\in\mathcal{S}$ and $\theta_{t,s}^0=0$ if $s\in\mathcal{C}$.

The state and the associated covariance are then given by
\begin{align}
\bm P_{t|t,s}^k&=(\bm \Gamma_{t|t,s}^k)^{-1}\\
\bm x_{t|t,s}^k&=\bm P_{t|t,s}^k\bm \gamma_{t|t,s}^k
\end{align}
With the updated state, the ($k+1$)th VB iteration can be carried out. The loop continues until when the number of VB iterations approaches $K$. For clarity, the resulting decentralized robust cubature information filter, labeled as dRCIF-1, is summarized in Algorithm~\ref{alg_df2}.

\subsection{A Reduced-Complexity Solution}

In this section, we propose a variant of the dRCIF-1, referred to as dRCIF-2 with reduced computational complexity and communication overhead.

In the dRCIF-1, consensus on likelihood is carried out in each VB iteration. Although this helps each sensor node use the information over the network (at least when $L$ is sufficiently large) to detect whether its local measurement is an outlier, the associated computational and communication costs may be excessive for applications involving, e.g., wireless sensor networks. In some cases, however, it is possible to reliably detect outliers by using only each sensor's own measurements~\cite{wang2018robust}. Hence, one possible way to reduce the computational and communication burden of the dRCIF-1 is to first perform VB iterations at each sensor node and then apply consensus on local likelihoods over the entire network. In this case, the local likelihood of each sensor node can be approximated by
\begin{align}
p(\bm y_{t,s}|\bm x_t,z_{t,s}^K)\propto\exp\left(-\frac{1}{2}(\bm x_t^T\bm I_{t,s}^K\bm x_t-2\bm x_t^T\bm i_{t,s}^K)\right)
\end{align}
where $\bm I_{t,s}^K$ and $\bm i_{t,s}^K$ are similarly defined as in~\eqref{I1} and~\eqref{i1}, i.e.,
\begin{align}
\bm I_{t,s}^K&= \langle z_{t,s}^K \rangle \bm H_{t,s}\bm R_{t,s}^{-1}\bm H_{t,s}\\
\bm i_{t,s}^K&=\langle z_{t,s}^K \rangle \bm H_{t,s}\bm R_{t,s}^{-1}\tilde{\bm y}_{t,s}
\end{align}

Similarly, for communication nodes, we have
\begin{align}
\bm I_{t,s}^K&=0\\
\bm i_{t,s}^K&=0
\end{align}
Then consensus on likelihood with $L$ iterations gives
\begin{align}
\bm I_{t,s}^{K,l}=\sum_{j\in N_s}\kappa_{s,j}\bm I_{t,j}^{K,l-1}\label{nv_c1}\\
\bm i_{t,s}^{K,l}=\sum_{j\in N_s}\kappa_{s,j}\bm i_{t,j}^{K,l-1}\label{nv_c2}
\end{align}
which are initialized by
\begin{align*}
&\bm I_{t,s}^{K,0}=\bm I_{t,s}^K,\ \bm i_{t,s}^{K,0}=\bm i_{t,s}^K.
\end{align*}
Finally, similar to the dRCIF-1, the correction step is implemented as
\begin{align}
\bm \Gamma_{t|t,s}^K&=\bm\Gamma_{t|t-1,s}^{K,L}+\delta_{t,s}\bm I_{t,s}^{K,L}\label{f_c1}\\
\bm \gamma_{t|t,s}^K&=\bm\gamma_{t|t-1,s}^{K,L}+\delta_{t,s}\bm i_{t,s}^{K,L}\label{f_c2}
\end{align}
where $\delta_{t,s}$ is the same scale parameter as defined in~\eqref{delta}. The state and the associated covariance are given by
\begin{align}
\bm P_{t|t,s}&=(\bm \Gamma_{t|t,s}^K)^{-1}\\
\hat{\bm x}_{t|t,s}&=\bm P_{t|t,s}^K\bm \gamma_{t|t,s}^K
\end{align}

The dRCIF-2 method is summarized in Algorithm~\ref{alg_df3}.

\begin{algorithm*}[!t]
\centering
\caption{Reduced-complexity decentralized robust CIF (dRCIF-2)} \label{alg_df3}
\begin{algorithmic}[0]
\State \textbf{Input:} $\mathcal{Y}_{1:T}$, $\hat{\bm x}_{0|0}$, $\bm P_{0|0}$, $\bm Q_{1:T}$, $\bm R_{1:T}$.
    \State \textbf{Output:} $\bm \hat{\bm x}_{t|t,s}$ and $\bm P_{t|t,s}$ for $t=1:T$ and $s\in\mathcal{D}$.
    \For {$t=1:T$}
    \State For each node $s\in\mathcal{D}$, compute $\{\hat{\bm x}_{t|t-1,s},{\bm P}_{t|t-1,s}\}$ and $\{{\bm \gamma}_{t|t-1,s},{\bm \Gamma}_{t|t-1,s}\}$ via \{\eqref{pu_1}--\eqref{iv_2}\}.
    \State For each node $s\in\mathcal{S}$, compute the pseudo-measurement matrix $\bm H_{t,s}$ via~\eqref{p_H}.
    \For {$l=1:L$}
    \State For each node $s\in\mathcal{D}$, consensus on prior information $\bm\gamma_{t|t-1,s}^l$ and $\bm \Gamma_{t|t-1,s}^l$ via \eqref{pr_c1} and \eqref{pr_c2};
    \EndFor
    \State For each node $s\in\mathcal{S}$, initialize $k=0$, $e_{t,s}^k$, $f_{t,s}^k$ and $\langle z_{t,s}^k\rangle=1$.
     \For {$k = 1 : K$}
     \State Calculate the local correction terms $\bm I_{t,s}^k$ and $\bm i_{t,s}^k$ via \eqref{I1} and \eqref{i1}with $\langle z_{t,s}^{k-1}\rangle$;
     \State Update the total information of the filtered state as $\bm\Gamma_{t|t,s}^k$ and $\bm \gamma_{t|t,s}^k$;
     \State Update the filtered state as $\bm P_{t|t,s}^k$ and $\hat{\bm x}_{t|t,s}^k$;
     \State Update $\langle z_{t,s}^{k}\rangle$ via \eqref{z} for $s\in\mathcal{S}$;
     \State Update $e_{t,s}^k$ and $f_{t,s}^k$ via \eqref{e} and \eqref{f} for $s\in\mathcal{S}$;
  \EndFor
  \State For each node $s\in\mathcal{S}$, set the local correction terms as $\bm I_{t,s}^{K,0}=\bm I_{t,s}^K$ and $\bm i_{t,s}^{K,0}=\bm i_{t,s}^K$;
  \State For each node $s\in\mathcal{C}$, set the local correction terms as $\bm I_{t,s}^{K,0}=0$ and $\bm i_{t,s}^{K,0}=0$;
  \For {$l=1:L$}
    \State For each node $s\in\mathcal{D}$, consensus on the local correction information $\bm I_{t,s}^{K,l}$ and $\bm i_{t,s}^{K,l}$ via \eqref{nv_c1} and \eqref{nv_c2};
  \EndFor
  \State For each node $s\in\mathcal{D}$, obtain the parameter $\kappa_t^s$ and update the total information $\bm \Gamma_{t|t,s}$ and $\bm\gamma_{t|t,s}$ via \eqref{f_c1} and \eqref{f_c2};
  \State ${\bm P}_{t|t,s}=(\bm \Gamma_{t|t,s})^{-1}$, $\hat{\bm x}_{t|t,s} = \bm P_{t|t,s}\bm \gamma_{t|t,s}$;
  \EndFor
  \end{algorithmic}
\end{algorithm*}

\begin{remark}
It is apparent that the consensus on prior step of both the dRCIF-1 and dRCIF-2 are the same. In this step, the quantities $\bm \Gamma_{t|t-1}$ and $\bm \gamma_{t|t-1}$ of each node are shared with its neighbors. $\bm \Gamma_{t|t-1}$ is a symmetric matrix with dimension $n\times n$, while $\bm \gamma_{t|t-1}$ is a vector with dimension $n\times 1$. Therefore, for the $j$-th node, it transmits $(n^2+3n)/2$ and receives $N_j(n^2+3n)/2$ real numbers in each consensus step.
\end{remark}

\begin{remark}
The main difference between the dRCIF-1 and dRCIF-2 is the way to implement the consensus on likelihood, which is carried out within the VB iterations in the dRCIF-1 while after the VB iterations in the dRCIF-2. In this step, the quantities $\bm I_t$ and $\bm i_t$ are shared, which have the same dimensions as these quantities in consensus on prior step. Therefore, there are $(n^2+3n)/2$ real numbers that are sent from $j$-th node to its neighbors in the dRCIF-2, while $K(n^2+3n)/2$ ($K$ is the number of the VB iterations) real numbers in the dRCIF-1.
\end{remark}

\begin{remark}
It is noted that while $\delta_t$ in~\eqref{delta} is calculated in a distributed approach, it is only dependent on the consensus weights (related to the structure of the network) and the consensus numbers. Therefore, for a static network (which is typical in many applications) it can be calculated offline, and incurs no communication overhead.
\end{remark}

\begin{remark}
The computational complexity of the VB iterations and the consensus on likelihood are, respectively, $\mathcal{O}(g_1(K))$ and $\mathcal{O}(g_2(L))$, where $g_1(\cdot)$ and $g_2(\cdot)$ are some functions with respect to their arguments. The computational complexity of the dRCIF-1 is approximately $\mathcal{O}(g_1(K))\mathcal{O}(g_2(L))$, while that of the dRCIF-2 is about $\mathcal{O}(g_1(K))+\mathcal{O}(g_2(L))$.
\end{remark}

\section{Application To Maneuvering Target Tracking}
\label{s5}

\begin{figure}[th]
\centering{\includegraphics[width=0.90\columnwidth]{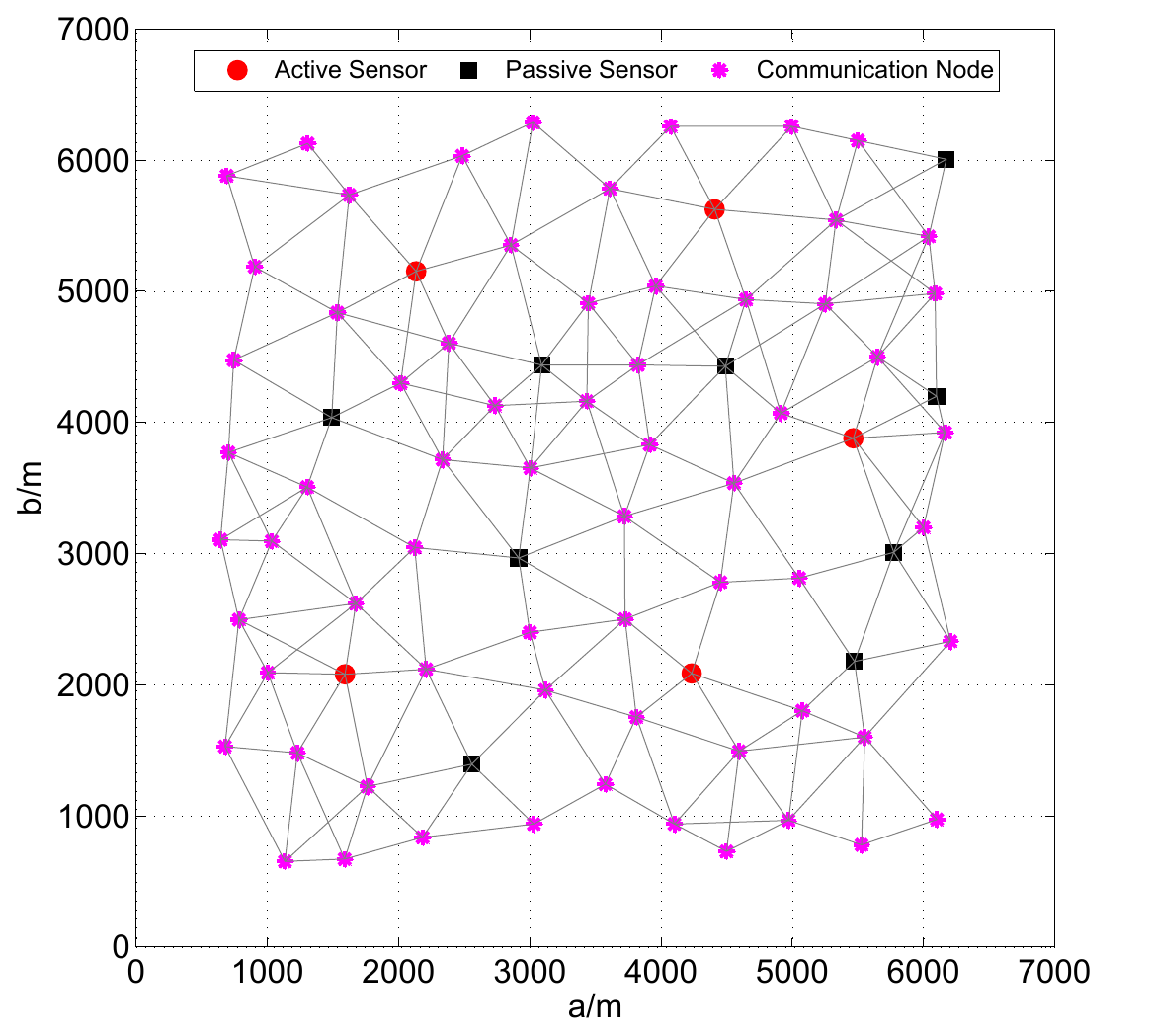}}
\caption{Graphical representation of the simulated network system}
\label{f1}
\end{figure}

In this section, we consider a target tracking problem to illustrate the performance of the proposed methods. A target maneuvers in an area which is surveilled by a networked sensing system. The networked system, as shown in Fig.~\ref{f1}, is equipped with $80$ nodes which include $5$ active sensors, $9$ passive sensors, and $66$ communication nodes. The presence of communication nodes is to enhance the connectivity of the networked surveillance system.

The dynamics of the moving target is described by a coordinated turning model with an unknown turning rate, i.e.,
\begin{align}
\bm x_{t+1}=\left(
\begin{array}{ccccc}
1&\frac{\sin(\omega_t\Delta t)}{\omega_t}&0&\frac{\cos(\omega_t\Delta t)-1}{\omega_t}&0\\
0&\cos(\omega_t\Delta t)&0&-\sin(\omega_t\Delta t)&0\\
0&\frac{1-\cos(\omega_t\Delta t)}{\omega_t}&1&\frac{\sin(\omega_t\Delta t)}{\omega_t}&0\\
0&\sin(\omega_t\Delta t)&0&\cos(\omega_t\Delta t)&0\\
0&0&0&0&1
\end{array}
\right)\bm x_t+\bm v_t
\label{ct}
\end{align}
where the state $\bm x_t$ is defined as $[a_t,\dot{a}_t,b_t,\dot{b}_t,\omega_t]^T$, containing the 2-D location $(a_t,b_t)$, the corresponding velocities $(\dot{a}_t,\dot{b}_t)$, and the turning rate $\omega_t$; $\Delta t=1$s is the sampling time, and $\bm v_t$ is a zero-mean Gaussian noise with covariance $\bm Q_t$:
\begin{align}
\bm Q_t=\left(\begin{array}{ccc}
q_1\bm M&0&0\\
0&q_1\bm M&0\\
0&0&q_2
\end{array}\right),
\bm M=\left(\begin{array}{cc}
\Delta t^3/3&\Delta t^2/2\\
\Delta t^2/2&\Delta t
\end{array}
\right)
\end{align}
where $q_1=0.1$ and $q_2=1.75\times 10^{-4}$. The trajectory of the moving target is randomly generated by~\eqref{ct} with the initial state $\bm x_0$ given by $\bm x_0 = [1000\text{ m},50\text{ m/s},2000\text{ m},-50\text{ m/s},0.053\text{ rad/s}]^T$. The initial condition of the state $\hat{\bm x}_{0|0}$ for each algorithm is chosen from a Gaussian $\mathcal{N}(\bm x_0,\bm P_0)$ with $\bm P_0=\text{diag}\left([10000,100,10000,100,3.04\times 10^{-6}]\right)$.

An active sensor provides the range and bearing measurements, given by
\begin{align}
\bm y_t^s=\left[\begin{array}{c}\sqrt{(a_t-p_x^s)^2+(b_t-p_y^s)^2}\\ \text{atan2}(b_t-p_y^s,a_t-p_x^s)\end{array}\right] + \bm w_t^s
\end{align}
where $(p_x^s,p_y^s)$ is the location of the active sensor; atan2 is the four-quadrant inverse tangent function and $\bm\omega_t^s$ is the measurement noise. Meanwhile a passive sensor measures bearing of the target,
\begin{align}
\bm y_t^s=\text{atan2}(b_t-p_y^s,a_t-p_x^s) + \bm w_t^s
\end{align}
We assume that the covariance of the nominal noise for the active sensor and passive sensor are, respectively, $\bm R_t=\text{diag}[10^2,1.22\times 10^{-5}]$ and $\bm R_t=1.22\times 10^{-5}$. In the simulation, the measurement noise is contaminated by outlier according the following model
\begin{align}
\bm w_t^s \sim\left\{
\begin{array}{ll}
\mathcal{N}(0,\bm R_t^s)&\text{with probability }1-\lambda\\
\mathcal{N}(0,\alpha\bm R_t^s)&\text{with probability } \lambda
\end{array}
\right.
\label{m_n}
\end{align}
where $\lambda$ and $\alpha$ are parameters to control the probability and power, respectively, of the outliers. This measurement model is a Gaussian mixture model, and has been widely used to evaluate the robustness of filtering in the presence of heavy-tailed measurement noises.

In the simulation, $M=100$ independent Monte Carlo runs are implemented and in each run the simulation length $T=50$. The root mean-square error (RMSE) of the target position, as well as the time-averaged RMSE (TRMSE), is employed as the performance metrics. For the centralized algorithms, the RMSE of position is defined as
\begin{align}
\text{RMSE}_t=\left(\frac{1}{M}\sum_{m=1}^M\left\|\bm p_t^{(m)}-\hat{\bm p}_t^{(m)}\right\|^2\right)^{1/2}
\end{align}
where $\bm p_t^{(m)}\triangleq(a_t^{(m)},b_t^{(m)})^T$ and $\hat{\bm p}_t^{(m)}\triangleq(\hat{a}_t^{(m)},\hat{b}_t^{(m)})^T$ are, respectively, the true and estimated position of the target at the $m$-th Monte Carlo run. For the decentralized methods, we employ the averaged RMSE, i.e.,
\begin{align}
\text{RMSE}_t=\left(\frac{1}{NM}\sum_{m=1}^M\sum_{s=1}^N\left\|\bm p_{t}^{(m)}-\hat{\bm p}_{t,s}^{(m)}\right\|_2^2\right)^{1/2}
\end{align}
where $\hat{\bm p}_{t,s}^{(m)}\triangleq(\hat{a}_{t,s}^{(m)},\hat{b}_{t,s}^{(m)})^T$ is the estimated target position of the $s$-th sensor. With the definition of the RMSE, the TRMSE of the position is given by
\begin{align}
\text{TRMSE}=\frac{1}{T}\sum_{t=1}^T\text{RMSE}_t
\end{align}

For comparison, we consider four existing filters:
\begin{inparaenum}[1)]
\item the clairvoyant centralized CIF which has the exact knowledge of the measurement noise model~\eqref{m_n}, denoted by cCIF-t;
\item the clairvoyant decentralized CIF with the exact knowledge of the measurement noise model~\eqref{m_n}, denoted by dCIF-t;
\item the robust decentralized CIF based on a student's t distribution\cite{dong2018robust}, denoted by dTCIF;
\item the interaction multiple model based robust decentralized CIF~\cite{battistelli2015consensus}, called dIMMCIF.
\end{inparaenum}
In the dTCIF, we set the parameters as recommended in~\cite{dong2018robust}. In the dIMMCIF, two models are employed based on~\eqref{m_n}, i.e.,
\begin{align*}
\bm w_t^s &\sim \mathcal{N}(0,\bm R_t^s)\quad\text{for the first model}\\
\bm w_t^s &\sim \mathcal{N}(0,\alpha\bm R_t^s)\quad\text{for the second model}
\end{align*}
The probability transition matrix for these two models is $[0.9,0.1;0.9,0.1]$ and the initial weights of these two model are 0.9 and 0.1, respectively.

\begin{figure}[th]
\centering{\includegraphics[width=\columnwidth]{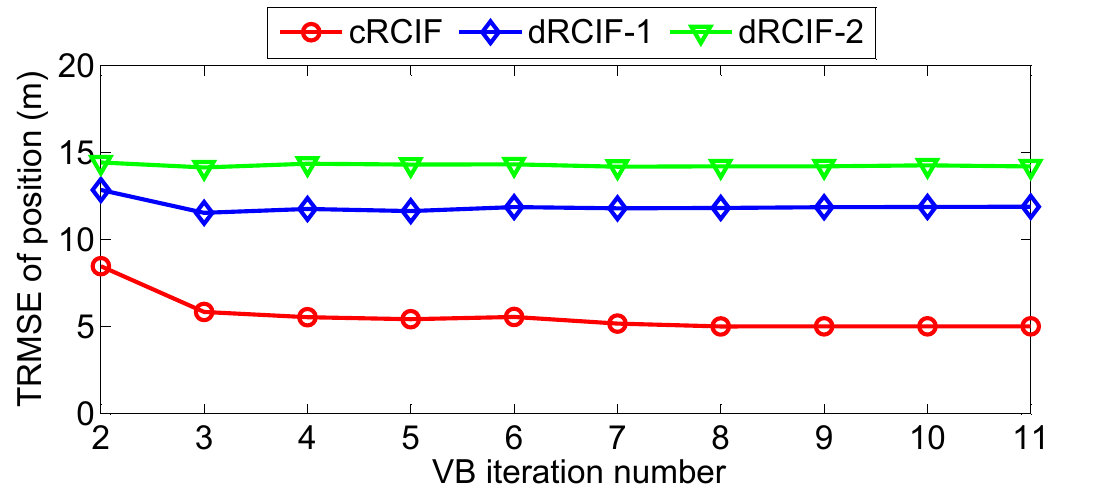}}
\caption{TRMSE of the position of different algorithms when $\lambda=0.4$ and $\alpha=100$ }
\label{f_iterloop}
\end{figure}

\begin{figure}[th]
\centering{\includegraphics[width=\columnwidth]{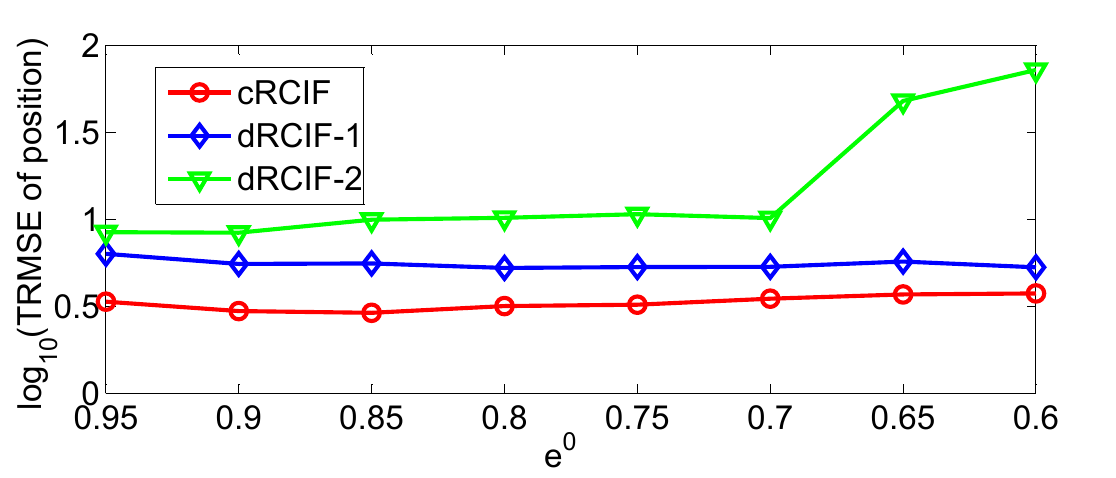}}
\caption{Logarithm of TRMSE of the position of different algorithms when $\lambda=0.1$ and $\alpha=100$ }
\label{f_inipara}
\end{figure}

First we evaluate how the iteration numbers of the VB and the initial parameters of the hierarchical model affect our proposed methods. Fig.~\ref{f_iterloop} shows the position TRMSEs of the proposed fusion algorithms versus the number of VB iterations when the initial parameters $e_{t,s}^0=0.9$ and $f_{t,s}^0=0.1$ in the scenario that $\lambda=0.4$ and $\alpha=100$. It is seen from the results that our methods achieve a stable estimate after two or three iterations. In the following, the default value of the VB iteration number of our methods is set to three. In Fig.~\ref{f_inipara}, we show the logarithm TRMSEs of position when $\lambda=0.1$ while $\alpha=100$, $e_{t,s}^0$ varies from $0.95$ to $0.6$ and $f_{t,s}^0$ equals $1-e_{t,s}^0$. It can be seen that both the cRCIF and dRCIF-1 are less sensitive to the initial value of $e_{t,s}^0$ and $f_{t,s}^0$ while the dRCIF-2 has a reasonable performance when $e_{t,s}^0$ is larger than $0.7$. This is because both the cRCIF and dRCIF-1 utilizes entire information for outlier detection while the dRCIF-2 only uses its own measurement. Due to the lack of adequate information, the prior probability of outlier occurrence plays an important role in outlier detection. In the following, the beta-Bernoulli parameters are set as $e_{t,s}^0=0.9$ and $f_{t,s}^0=0.1$ so that $e_{t,s}^0/(e_{t,s}^0+f_{t,s}^0)$ is close to $1$. Although this setting treats the measurement as a nominal one with a high probability, the VB outlier detection procedure, as illustrated in the following simulations, is able to distinguish between the outlier and nominal measurement.

Table~\ref{table} provides the averaged position TRMSE of five decentralized solutions with different consensus steps. Since the dIMMCIF is based on the CM strategy, more consensus steps are implemented to obtain a reasonable estimate. It can be seen that the performance of the decentralized solutions, as expected, improves as the consensus step increases. Among all decentralized fusion algorithms, our proposed dRCIF-1 has the smallest gap compared with the benchmark solution dCIF-t, followed by the dRCIF-2 and dTCIF, which are similar. Even though the dIMMCIF has more consensus steps, its performance is still the worst. As mentioned before, the consensus step is closely related to the computational complexity and communication overhead, in the following we set $L=10$ for the dIMMCIF while $L=5$ for the other four.

\begin{table}[!h]
\centering
\caption{TRMSE of position for different algorithms with different consensus steps}
\label{table}
\begin{tabular}{cccccc}
\hline
        & $L=1 $  & $L=2$   & $L=3 $  & $L=4$   & $L=5$   \\
\hline
dRCIF-1 & 15.49 & 10.70 & 8.97  & 7.87  & 7.41  \\
dRCIF-2 & 15.79 & 11.69 & 10.45 & 10.11 & 9.63  \\
dTCIF   & 26.36 & 15.14 & 12.73 & 11.63 & 10.24 \\
dCIF-t  & 12.21 & 8.73  & 7.69  & 7.26  & 6.90  \\
\hline
        & $L=6$   & $L=7$   & $L=8$   & $L=9$   & $L=10$   \\
\hline
dIMMCIF & 21.87 & 21.89 & 21.26 & 20.43 & 19.49 \\
\hline
\end{tabular}
\end{table}

\begin{figure}[th]
\centering{\includegraphics[width=0.9\columnwidth]{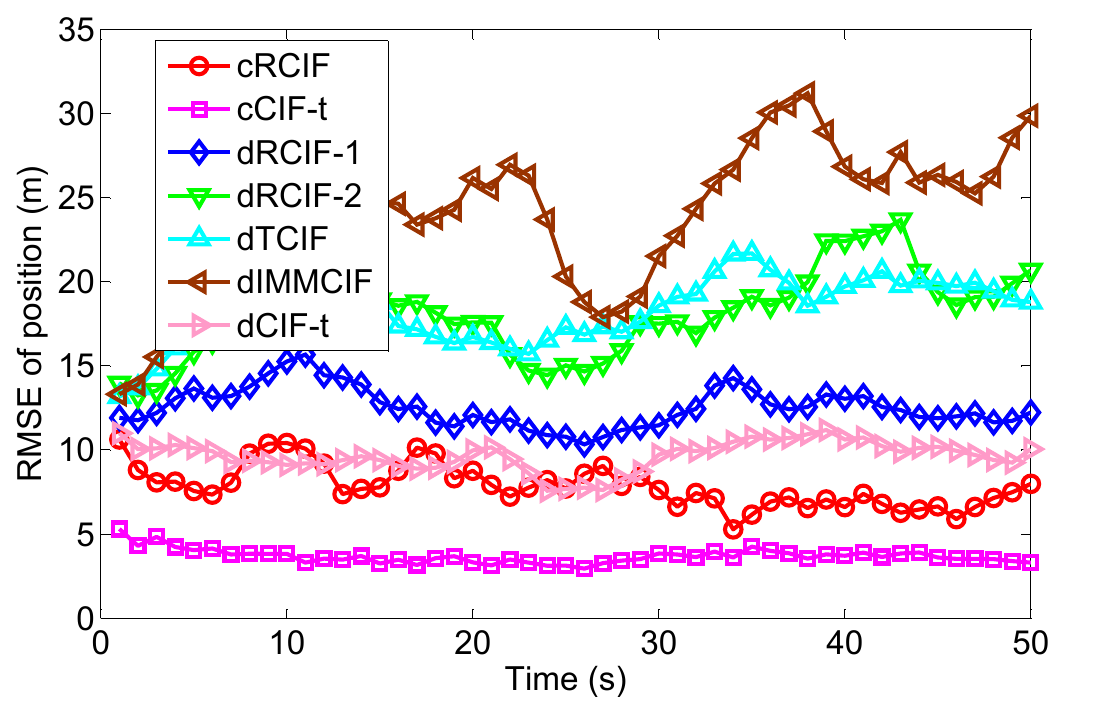}}
\caption{RMSE of the position of different algorithms when $\lambda=0.5$ and $\alpha=100$ }
\label{f_rmse}
\end{figure}

\begin{figure}[th]
\centering
{\includegraphics[width=0.9\columnwidth]{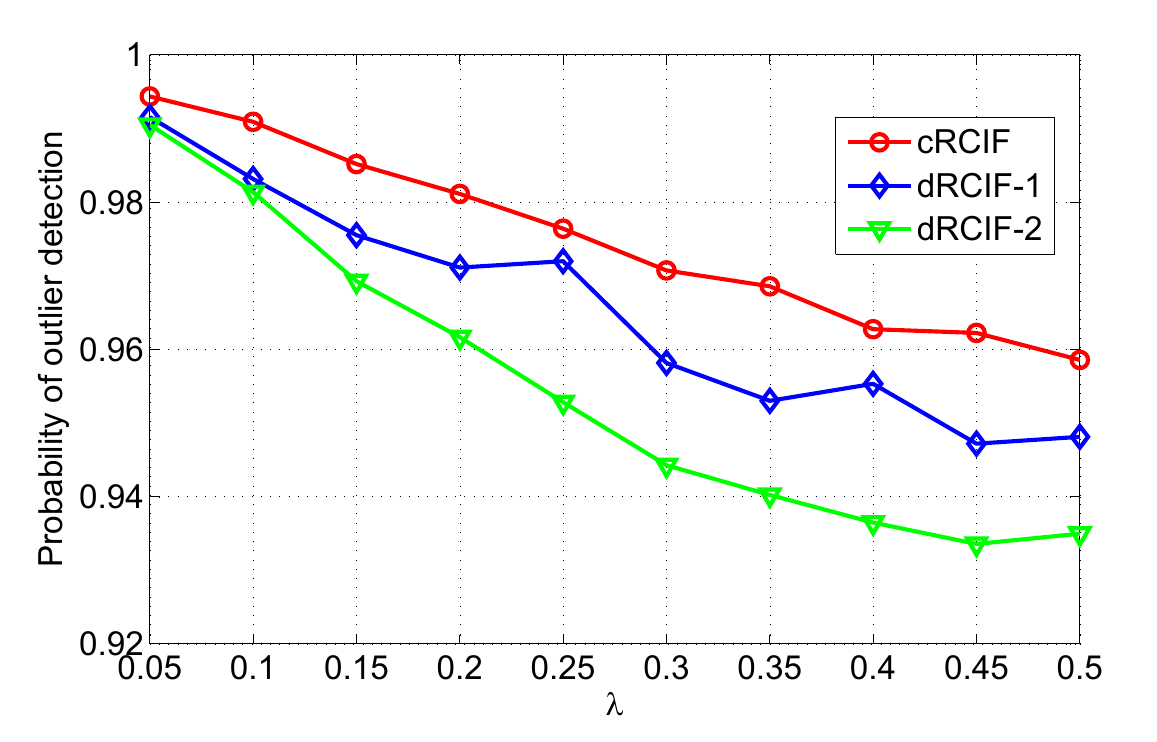}}
\caption{The probability of outlier detection of the proposed method when $\lambda$ varies and $\alpha=500$ }
\label{f_od_pr}
\end{figure}

Fig.~\ref{f_rmse} shows the RMSEs of the position when $\lambda=0.5$ and $\alpha=100$. Among all decentralized solutions, the performance of the proposed dRCIF-1 is closest to that of the benchmark dCIF-t. The computationally simpler dRCIF-2 shares a similar performance as that of the dTCIF, and both are better than the dIMMCIF. The centralized benchmark solution cCIF-t provides an overall smallest RMSE, and the proposed centralized solution cRCIF performs somewhat better than the decentralized benchmark, i.e., the dCIF-t. Fig.~\ref{f_od_pr} illustrates the outlier identification ability of the proposed algorithms under different $\lambda$ when $\alpha$ is set to $500$. There is no doubting that the centralized method is superior the other two decentralized solutions. The dRCIF-1 outperforms the dRCIF-2, and the major reason is that the outlier-detection procedure in the dRCIF-1 is within the consensus iterations.

\begin{figure}[th]
\centering{\includegraphics[width=0.9\columnwidth]{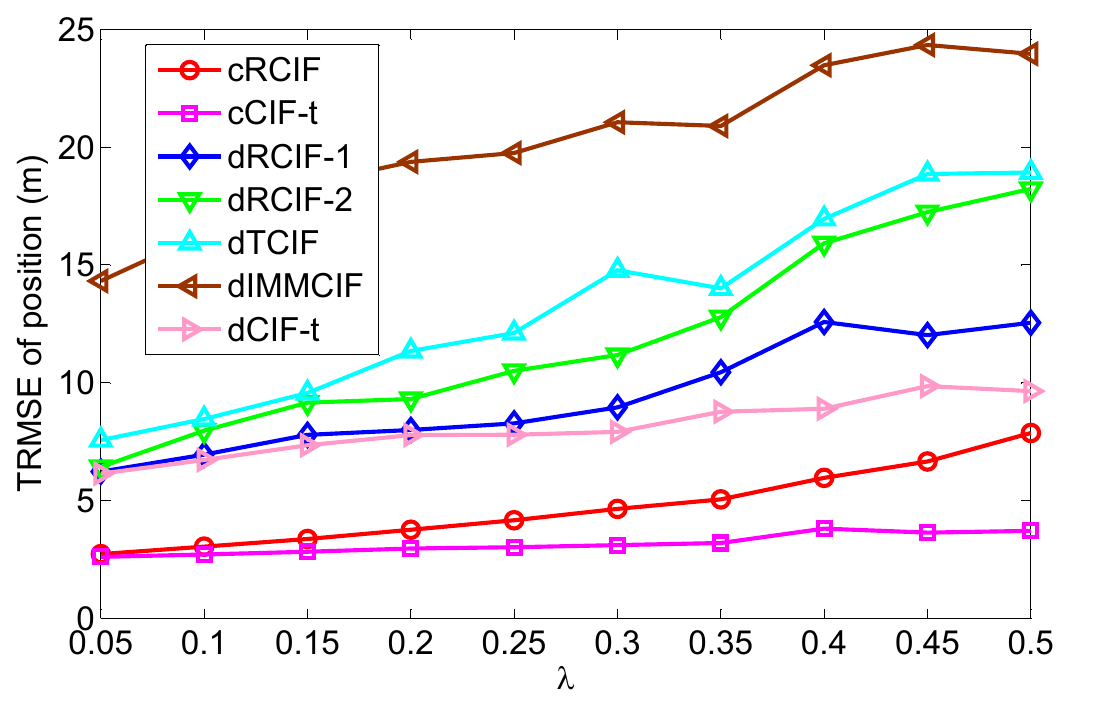}}
\caption{TRMSEs of position for the different algorithms with varying $\lambda$ when $\alpha=100$}
\label{f_varlambda}
\end{figure}

\begin{figure}[th]
\centering{\includegraphics[width=0.9\columnwidth]{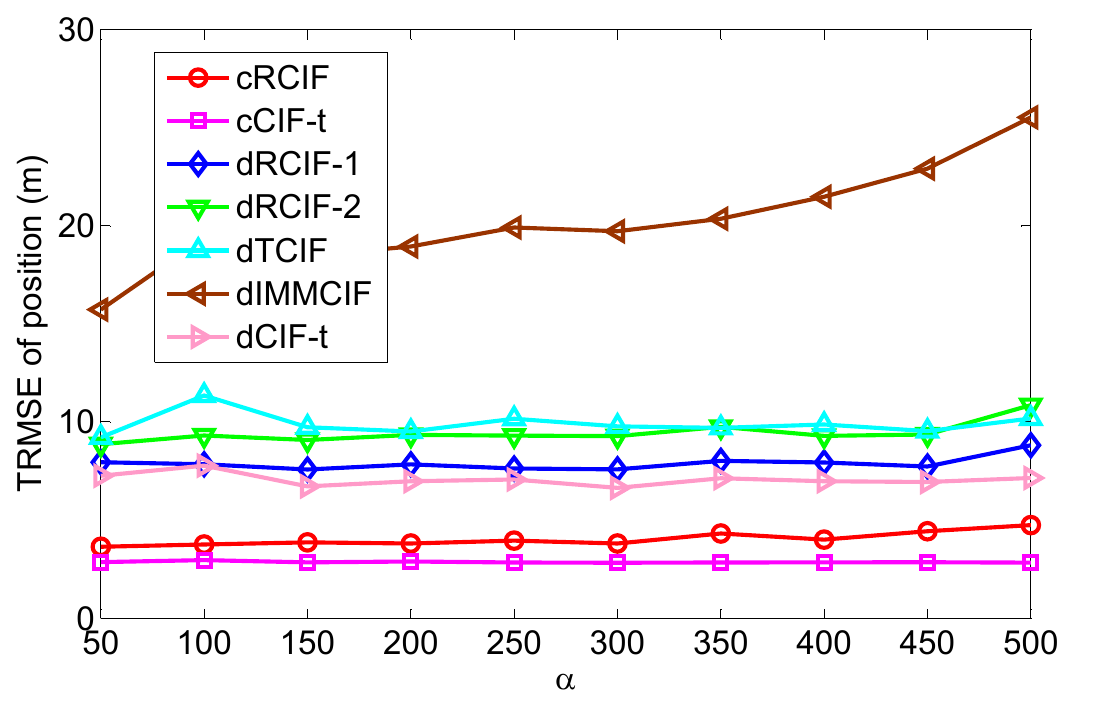}}
\caption{TRMSEs of position for the different algorithms with varying $\alpha$ when $\lambda = 0.2$}
\label{f_varalpha}
\end{figure}

Finally we examine how the contamination ratio and the power of the contaminating noise influence the proposed solutions. Fig.~\ref{f_varlambda} plots the position TRMSEs of the various information fusion algorithms when $\alpha=100$ and $\lambda$ varies from $0.05$ to $0.5$, while in Fig.~\ref{f_varalpha} we show the similar results with varying $\alpha$ and $\lambda$ fixed at $0.2$. From these two figures, we can see that the TRMSEs of all algorithms increase along with $\lambda$, while all except the dIMMCIF (due to the fact that two models are involved) are nearly unaffected by the growth of $\alpha$. This shows that these algorithms are sensitive to the contamination ratio while less sensitive to the power of the contaminating noise.

\section{Conclusion}
\label{s6}

In this paper, we considered the information fusion problem of networked systems where measurements may be disturbed by outliers. We introduced a hierarchical measurement model to take potential outliers into consideration. Specifically, we utilized a binary variable, which has a beta-Bernoulli prior, for each measurement to indicate whether it is a nominal observation or an outlier. Based on the proposed outlier-detection measurement model, we first developed a centralized robust information fusion algorithm, which jointly infers the state and indicator variable via a variational Bayesian method. Furthermore, we proposed two decentralized robust solutions by integrating the HCICM consensus strategy with outlier detection and inference. Simulation results illustrated that the proposed approaches can achieve better performances compared the existing ones with outlier contaminated measurements.

\appendices

\numberwithin{equation}{section}
\setcounter{equation}{0}

\section{Cubature Information Filter}

Consider the SSM model described in~\eqref{process} and~\eqref{measure} with only one sensor, the conventional CIF is briefly summarized as follows for easy reference.

\subsubsection{Initialization}

We initialize the CIF with $\bm x_0\thicksim\mathcal{N}(\hat {\bm x}_{0|0},\bm P_{0|0})$, and generate the basic weighted cubature point set, i.e., $\{\bm \eta_{i},\omega_i\}$ for $i=1,\cdots,2n$, where $\omega_i=1/(2n)$ and $\bm \eta_{i}=\sqrt{n}[\bm \Pi]_i$. Here $[\bm \Pi]_i$ denotes the $i$th column of $\bm \Pi \in\Re^{n\times (2n)}$ which is a block matrix given by $\bm \Pi =[\bm I\ -\bm I]$ with $\bm I$ being the identify matrix.

\subsubsection{Prediction}

Assume that at the time instant $(t-1)$, the posterior distribution of state ${\bm x}_{t-1}$ is approximated by $\mathcal{N}(\hat{\bm x}_{t-1|t-1},\bm P_{t-1|t-1})$. The transformed sigma-points and their associated weights related to $\mathcal{N}(\hat{\bm x}_{t-1|t-1},\bm P_{t-1|t-1})$ are generated as:
\begin{align}
\bm P_{t-1|t-1}&=\bm S_{t-1|t-1}\bm S_{t-1|t-1}^T\label{ckf_be}\\
\bm\eta_{i,t-1}&=\bm S_{t-1|t-1}\bm\eta_i+\hat{\bm x}_{t-1|t-1}
\end{align}
And then the predicted state and its associated covariance are updated by
\begin{align}
\bm \chi_{i,t-1}=&f(\bm \eta_{i,t-1})\\
\bm \hat{\bm x}_{t|t-1}=&\sum_{i=1}^{2n}\omega_i\bm  \chi_{i,t-1}\label{pu_1}\\
\bm P_{t|t-1}=&\sum_{i=1}^{2n}\omega_i(\bm \chi_{i,t-1}-\bm \hat{\bm x}_{t|t-1})\notag\\
&\qquad\quad\times(\bm \chi_{i,t-1}-\bm \hat{\bm x}_{t|t-1})^T+\bm Q_{t-1}\label{pu_2}
\end{align}
The prior information of the state is then written as
\begin{align}
\bm \Gamma_{t|t-1}&=\bm P_{t|t-1}^{-1}\label{iv_1}\\
\bm \gamma_{t|t-1}&=\bm \Gamma_{t|t-1}\bm \hat{\bm x}_{t|t-1}\label{iv_2}
\end{align}

\subsubsection{Filtering}

Using the statistical linear error propagation methodology~\cite{lee2008nonlinear}, the pseudo-measurement matrix is defined as
\begin{align}
\bm H_{t}=\bm\Gamma_{t|t-1}\bm P_{xy}\label{p_H}
\end{align}
where $\bm P_{xy}$ is the cross covariance calculated by
\begin{align}
\bm P_{t|t-1}&=\bm S_{t|t-1}\bm S_{t|t-1}^T\\
\bm \varrho_{i,t}&=\bm S_{t|t-1}\bm\eta_i+\bm \hat{\bm x}_{t|t-1}\\
\bm{\hat y}_t &= \sum_{i=1}^{2n}\omega_i\bm h(\varrho_{i,t})\\
\bm P_{xy}&=\sum_{i=1}^{2n}\omega_i\left(\bm \varrho_{i,t}-\bm{\hat x}_{t|t-1}\right)\left(\bm \zeta_{i,t}-\bm{\hat y}_t\right)^T
\end{align}
With $\bm H_{t}$, the correction information terms are given by
\begin{align}
\tilde{\bm y}_t&=\left(\bm y_t-\hat{\bm y}_t+\bm H_t\hat{\bm x}_{t|t-1}\right)\label{y_w}\\
\bm I_t&=\bm H_{t}\bm R_t^{-1}\bm H_{t}^T\label{up_iv1}\\
\bm i_t& = \bm H_{t}\bm R_t^{-1}\tilde{\bm y}_t\label{up_iv2}
\end{align}
Finally, the information formate of the filtered state is given by
\begin{align}
\bm\Gamma_{t|t}&=\bm \Gamma_{t|t-1}+\bm I_t\\
\bm \gamma_{t|t}&=\bm \gamma_{t|t-1}+\bm i_t
\end{align}
and the filtered state is recovered as
\begin{align}
\bm P_{t|t}&=\bm \Gamma_{t|t}^{-1}\\
\hat{\bm x}_{t|t}&=\bm P_{t|t}\bm \gamma_{t|t}
\end{align}


\end{document}